# Evidence For Quasi-Periodic Oscillations In The Recurrent Bursts From SGR 1806-20

Ahmed M. El-Mezeini[*][†] and Alaa I. Ibrahim[*][‡][¶]

[*]*Department of Physics, The American University in Cairo, New Cairo, 11835, Egypt*
[†]*Department of Applied Mathematics and Theoretical Physics, Centre of Mathematical Sciences, University of Cambridge, Wilberforce Road, Cambridge CB3 0WA, United Kingdom*
[‡]*Kavli Institute for Astrophysics & Space Research, Massachusetts Institute of Technology, MA 02139, USA*
[¶]*Department of Physics, Faculty of Science, Cairo University, Giza, 12613, Egypt*

**Abstract.** We present evidence for Quasi Periodic Oscillations (QPOs) in the recurrent outburst activity from SGR 1806-20 using Rossi X-ray Timing Explorer (RXTE) observations during November 1996. Searching for QPOs in a sample of 30 bursts at similar frequencies to those previously reported in the December 27, 2004 giant flare, we find evidence for a QPO in a burst at 648 Hz at 5.17σ confidence level, lying within 3.75% from the 625 Hz QPO discovered in the giant flare. Two additional features are also detected around 84 and 103 Hz in two other bursts at 4.2σ and 4.8σ confidence level, respectively, which lie within 8.85% and 11.83% respectively from the QPO at 92.5 Hz also detected in the giant flare. Accounting for the number of bursts analyzed the confidence levels for the 84, 103 and 648 Hz becomes 3σ, 3.6σ and 3.4σ respectively. Extending our search to other frequency ranges, we find candidates at 1096, 1230, 2785 and 3690 Hz in 3 different bursts with confidence levels lying between 4.14σ-4.46σ, which is reduced to 2.3σ-3σ after accounting for a certain confirmation bias in each case. The fact that we can find evidence for QPOs in the recurrent bursts at frequencies relatively close to those found in the giant flare is intriguing. We examine the candidate QPOs in relation with those found in the giant flare and discuss their possible physical origin.

**Keywords:** pulsars: individual (SGR 1806-20)—stars: flare—stars: neutron—stars: oscillations— X-rays: bursts.

## INTRODUCTION

Soft gamma repeaters (SGRs) are modeled as magnetars, which are neutron stars, characterized by intense magnetic field of the order $10^{14}$ - $10^{15}$ Gauss [1, 2]. SGRs emit short and recurrent bursts of soft γ-rays and are considered as sources of high-energy transient bursts that were later also found to be persistent X-ray pulsars with periods of several seconds spinning down rapidly. Until recently there were only four confirmed SGRs, three are in our Galaxy and one in the Large Magellanic Cloud [3]. Recently two additional SGRs were discovered which brings the total number of confirmed SGRs to six. The well-known behavior of SGRs is their repetitive emission of bright bursts of low-energy γ-rays with short durations (t ~ 0.1 s) and peak luminosities reaching up to $10^{41}$ erg s$^{-1}$ well above the classical Eddington luminosity $L_{edd} = 2\times10^{38}$ erg s$^{-1}$ for a 1.4 $M_\circ$ neutron star. Less frequently the highly magnetized neutron stars undergo enormously energetic events known as giant flares with peak luminosities of the order $10^{44}$ - $10^{46}$ erg s$^{-1}$. The magnetic instability causes the eruption of these giant flares accompanied by large-scale fracturing of the crust of the neutron star, which would excite toroidal vibrational modes in the neutron star [2, 4-6]. Global toroidal modes excited by crust fracturing are regularly observed by Earth seismologists in earthquakes [7].

The magnetar model proposes that the dominant source of energy of SGRs is their intense magnetic field, where the short bursts are created as a result of the outward propagation of Alfvén waves through the magnetosphere driven by magnetic field diffusion through small cracks in the neutron star crust. The highly energetic giant flares are initiated by the sudden catastrophic reconfiguration of the intense magnetic field. It has been suggested that the coupling between the magnetic field and the charged particles in the neutron star would trigger star quakes causing global fracturing in the neutron star crust and result in the modulation of the X-ray light curve. These oscillations associated with toroidal vibrational modes of the neutron star are easier to excite than other possible oscillation mechanisms [8]. The frequency harmonics modes excited depend on the neutron star mass and radius, crustal composition and magnetic field. Detection of these toroidal crustal modes has a great potential of revealing and probing the stellar structure, composition, magnetic field configuration and testing the neutron star equation of state [9-15].

On December 27, 2004 SGR 1806-20 emitted the most powerful flare ever recorded from a magnetar which was detected by a number of different satellites where a peak luminosity of the order $10^{46}$ erg s$^{-1}$ [16]. In a pioneering paper, the detection of QPOs in the 18-93 Hz range in the decaying tail of the flare was reported using RXTE/PCA data [17]. Further analysis of the giant flare using observations from Ramaty High Energy Solar Spectroscopic Imager (RHESSI), a satellite that covers a wider energy band than RXTE confirmed the detection of the 18 and 92.5 Hz QPOs and reported the additional presence of a broad 26 Hz QPO and high frequency QPOs at 626.5 and 1840 Hz [18]. Prompted by the SGR 1806-20 results, the August 27, 1998 giant flare from SGR 1900+14 was reanalyzed using RXTE/PCA observations. This analysis revealed 4 QPOs in the range 28-155 Hz [19]. In both cases preliminary identifications can be made with a sequence of toroidal modes. Motivated by these results, we address here the question of whether similar QPOs may be detected in the typical SGR outbursts. We analyzed the November 1996 outburst from SGR 1806-20 using RXTE/PCA observations to search for similar QPOs or ones at different frequencies. We then compare our results with previously reported QPOs in SGR 1806-20 and SGR 1900+14 and discuss the implications of these results.

## DATA ANALYSIS AND RESULTS

The discovery of QPOs in giant flares stimulated us to analyze the November 1996 outburst from SGR 1806-20 using RXTE/PCA observations initially in search for QPOs at similar frequencies as those previously reported and for analogous phenomenology at other frequencies.

During the November 1996 outburst from SGR 1806-20, RXTE captured several hours' worth of data where all 5 proportional counter units (PCU) of the proportional counter array (PCA) were fully functional and in operation. The RXTE/PCA detects X-ray activities and emissions in the energy range 2-60 KeV with minimum time resolution 1 µs. Initially, we searched for QPOs around the same frequency range as the previously reported QPOs using the full energy range. The search was carried out on seven event mode files that contained several bursts where we generated a 16 ms light curve for the whole event. Afterwards, the bursts light curves were re-binned using a 125 µs time resolution (Nyquist frequency ~ 4 kHz) for the power spectrum generation in search of any significant features in the power spectrum.

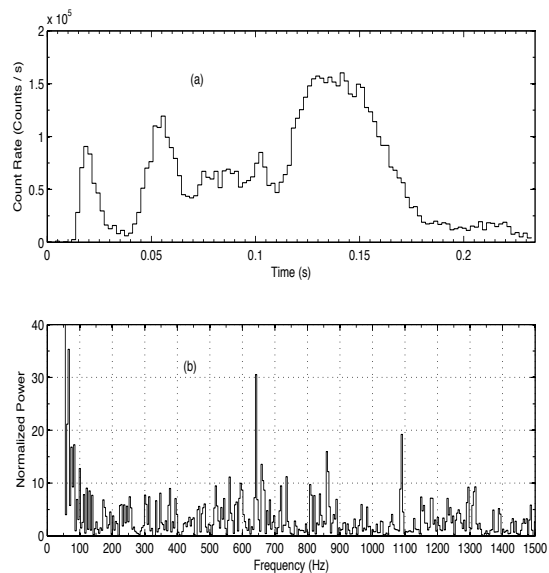

**FIGURE 1.** (a) Burst light curve starting at 04:41:53.480 UTC sampled at 2 ms. (b) The Power spectrum of the burst at a frequency resolution Δν= 4.27 Hz clearly shows two power peaks at 648 and 1096 Hz.

Fourier analysis was applied to generate the power spectrum using FFT algorithm and the power spectrum was normalized using Leahy normalization method [20, 21], where we considered the local noise level to be an average for the power level before and after the QPO peak. We first discuss the QPOs that lie at frequencies similar to those previously detected. In one intense burst at 04:41:53.480 UTC and lasting for 0.25 s, two peaks in the power spectrum around 648 and 1096 Hz were detected at a frequency resolution Δν= 4.27 Hz as shown in Fig. 1. Using Lorentzian fitting model to fit the first peak yields centroid frequency of 648.5 ± 0.15 Hz and a half width at half maximum (HWHM) $\sigma_\nu$ = 2.51 ± 0.09 Hz, corresponding to a coherence value $Q \equiv \nu_o/2\sigma_\nu$ of 130. The null hypothesis test was formed to estimate the Null Probability significance using a $\chi^2$ distribution resulted in a single trial probability of 2.33×10$^{-7}$ corresponding to a confidence level interval of 5.17 σ, which we consider quite robust.

Figure 2a shows an interesting feature that was detected in the power spectrum at frequency resolution Δν= 11.38 Hz for a relatively short burst starting at 08:57:02.083 UTC. Again, we use Lorentzian fitting model for this feature yields a centroid frequency of 103.44 ± 0.43 Hz with a coherence value Q ~ 4.86 and forming the null hypothesis test yields a null probability of $1.43 \times 10^{-6}$ corresponding to a confidence level of 4.82 σ which is also quite robust. We also report a less significant yet intriguing QPO candidate in the power spectrum of the burst shown in Fig. 2c having a centroid frequency 84.31 ± 0.66 Hz and Q ~ 4.84 with single trial null probability $3.4 \times 10^{-5}$ corresponding to significance level of 4.2 σ. The 84 Hz feature coincides with the QPO around 84 Hz previously reported in SGR 1900+14 giant flare on August 1998 [19].

06:18:44.446 UTC plotted at 2 ms time resolution. (d) The average power spectrum at 11.38 Hz resolution of the burst in panel c reveals a peak at 84 Hz.

High frequency features were also detected with the peaks centered on the frequencies 1096, 1230, 2785 and 3690 Hz. From the power spectrum shown in Fig. 1 we see a peak at a frequency higher than 1 kHz, which is energy dependent and not visible in the 2-10 KeV energy range. Using Lorentzian fitting model yields a centroid frequency of 1095.88 ± 0.21 Hz and a HWHM of 2.88 ± 0.17 Hz corresponding to a coherence value Q ~ 190. We estimate the single trial null probability to be $1.37 \times 10^{-5}$. It is obvious from Fig. 3 showing the power spectrum of one of the sporadic bursts starting at 06:18:44.446 UTC that we have two peaks around 1230 and 3690 Hz, where we apply the Lorentzian fitting model yields centroid frequencies 1229.80 ± 0.80 and 3690.74 ± 0.67 Hz respectively. Forming the null hypothesis results in single trial significance of 4.14 σ for each feature. One final high frequency QPO candidate around 2785 Hz is visible in the power spectrum of an approximately 0.6 second long intense burst starting around 12:18:01.487 UTC. as seen in Fig. 4 which is obtained from fitting the narrow peak using Lorentzian model gives a QPO candidate at 2785.36 ± 0.20 Hz and an HWHM of 0.75 ± 0.19 Hz with single trial probability of $7.97 \times 10^{-6}$.

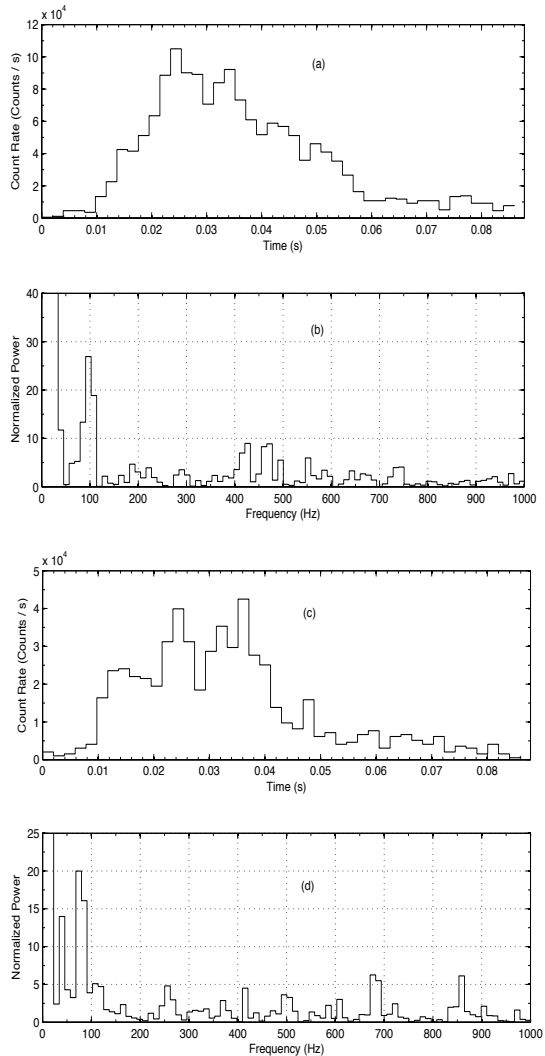

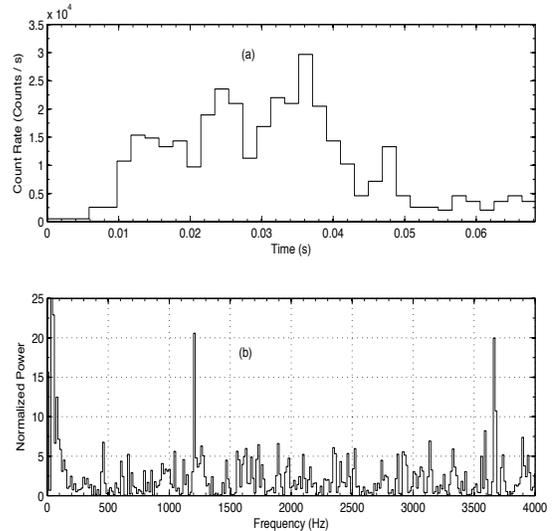

FIGURE 2. (a) The Nov. 18 X-ray burst light curve starting at 08:57:02.083 UTC plotted at 2 ms time resolution. (b) The Power spectrum of the burst in panel a reveals a peak at 105 Hz. (c) The Nov. 18 X-ray burst light curve starting at

FIGURE 3. (a) The Nov. 18 X-ray burst light curve with 2 ms time resolution starting at 06:18:44.134 UTC. (b) The Power spectrum of the burst clearly shows two power peaks at 1230 and 3690 Hz.

Our results show that the detected QPO candidate around 648 Hz is within 3.75% of the 626.5 Hz feature reported in the analysis of SGR 1806-20 giant flare using RHESSI observations while the 104.8 Hz feature lies within 11.83% of the previously detected feature at 92 Hz and the 84 Hz QPO candidate lies within 8.85% of the 92 Hz QPO and coincides with the 84 Hz QPO discovered in the SGR 1900+14 giant flare. We conjecture that for these three cases we do not need to account for the confirmation bias given that we already knew that a QPO has been detected in this frequency range previously for the same source SGR 1806-20 hence we account only for the number of bursts we investigated which brings the confidence level for the 84, 103 and 648 Hz QPOs to ~ 3σ, 3.6σ and 3.4σ respectively. For the high frequency QPOs 1096, 1230, 2785 and 3690 Hz we have no prior knowledge of these QPO frequencies so we need to account for the number of frequency bins searched and the total number of bursts analyzed which reduces the confidence level to the 2.3σ-3σ range for the high frequency QPOs. Further analysis led to the generation of light curves and power spectrums using energy windowing in the 2-10, 2-20 and 2-60 KeV energy ranges to examine the variation of in the light curve and power spectrum and verify the existence and magnitude of QPO candidates in different energy ranges. All QPO candidates we report seemed to be consistent in all energy ranges except for the 1096 Hz QPO candidate, which was no longer visible in the 2-10 KeV energy range. In our analysis the reported significances for the above QPOs were calculated in the 2-20 KeV energy range where most of the arriving photons are located as we noticed from the photon energy distribution.

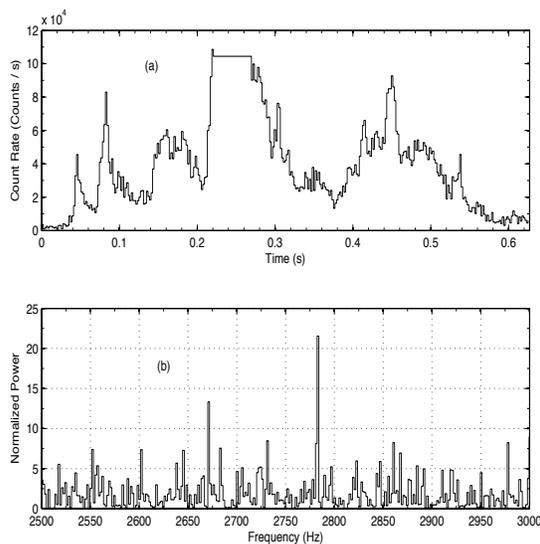

**FIGURE 4.** (a) The Nov. 18 1996 X-ray burst light curve starting at 12:18:01.487 UTC. (b) The Power spectrum of the burst clearly shows a power peaks at 2785 Hz.

We also investigated the intrinsic variability in the multi-peaked light curves where this periodicity could introduce peaks in the power spectrum in 10-100 Hz in the light curve and power spectrum in Fig. (1) and 2.5-100 Hz in the light curve and power spectrum in Fig. (4). However none of these frequency ranges overlap with QPO candidates detected in the power spectrum for both cases, which implies that these QPO candidates are not a result of the intrinsic variability or periodicity in the light curve.

## DISCUSSION

In a landmark paper Quasi - periodic oscillations were discovered in SGR 1806-20 during the giant flare on December 27, 2004 using RXTE/PCA observations at different frequencies including the detection of QPOs at 92.5 Hz which was seen only over phases of the 7.56 second spin period yet no energy dependence was detected. QPOs at 18 and 30 Hz were also detected however no specific pulse phase dependence was obvious [17]. Shortly after the detection of QPOs in the SGR 1806-20, detailed timing analysis of the August 27, 1998 giant flare from SGR 1900+14 using RXTE/PCA yielded the discovery and detection of QPOs at 84 Hz which appears to be associated with a particular rotational phase and two other QPOs were strongly detected at 53.5 Hz and 155.1 Hz and another broad feature at 28 Hz with lower significance. Our analysis of the high-energy emission bursts from SGR 1806-20 shown above show interesting similarities to the results reported earlier [17, 18, 19]. The properties of the X-ray bursts that revealed the existence of QPO candidates are summarized in Table 1.

**TABLE 1.** Properties of X-ray Bursts and QPO Candidates

| Event Date and Start Time (UTC) | Centroid Frequency (Hz) | Single Trial Confidence Level Interval | Coherence Value |
|---|---|---|---|
| Nov. 18, 1996 at 04:41:53.480 | 648 & 1096 | 5.17σ & 4.35σ | 130 & 190 |
| Nov. 18, 1996 at 06:18:44.134 | 1230 & 3690 | 4.14σ for both | 77 & 220 |
| Nov. 18, 1996 at 06:18:44.446 | 84 | 4.2 σ | 5 |
| Nov. 18, 1996 at 08:57:02.083 | 104 | 4.82σ | 5 |
| Nov. 18, 1996 at 12:18:01.487 | 2785 | 4.46σ | 1860 |

We found three intriguing QPO candidates in SGR 1806-20 at 84, 104, and 648 Hz. The 84 Hz QPO candidate matches the feature found in SGR 1900+14 and lies within 8.85% of the 92.5 Hz QPO while the energy dependent 105 Hz feature lies within 11.83% the 92.5 Hz feature in from the same source. The high frequency QPO candidate at 648 Hz is within 3.75% of the 625 Hz QPO discovered in 1806. The similarity in the detected QPOs is quite intriguing in its own right and may imply an analogous underlying physical mechanism. It was conjectured that toroidal oscillations of the neutron star might explain the observed frequencies [17]. We discuss our findings in context of this model.

It has been estimated for a non-rotating, non-magnetic neutron star [9], the Eigen frequency of the fundamental toroidal mode $l=2$, $n=0$ denoted $_2\nu_o$ to be

$$\nu(_2t_o) = 29.8 \frac{\sqrt{1.71 - 0.71 M_{1.4} R_{10}^{-2}}}{0.87 R_{10} + 0.13 M_{1.4} R_{10}^{-1}} \quad Hz \quad (1)$$

Where $R_{10} \equiv R/10$ km and $M_{1.4} \equiv M/1.4 M_\odot$. If a twisted magnetic field is embedded throughout the neutron star crust, the Eigen frequencies need to be modified to

$$\nu \approx \nu_o \left[ 1 + \left( \frac{B}{B_\mu} \right)^2 \right]^{1/2} \quad Hz \quad (2)$$

Where $\nu_o$ is the non-magnetic Eigen frequency, as in Eq. (1) for the fundamental mode $\nu(_2t_0)$. The higher order mode Eigen frequencies were shown to be scaled as $\nu(_lt_o) \propto [l(l+1)]^{1/2}$ [10]. The Eigen frequencies of the higher order $n=0$ toroidal modes were derived by Strohmayer and Watts under the assumption that the magnetic tension boosts the field isotropically. One must note that the degree to which the magnetic field modifies the Eigen frequency is highly dependent on the magnetic field configuration and other non-isotropic effects, which could alter this correction significantly [12]. The derived Eigen frequencies of the higher order toroidal modes ($l>0$ and $n=0$) under the assumption of isotropic magnetic field are given by

$$(3)$$

$$\nu(_lt_o) = \nu(_2t_o) \left[ \frac{l(l+1)}{6} \right]^{1/2} \left[ 1 + \left( \frac{B}{B_\mu} \right)^2 \right]^{1/2} \quad Hz$$

Where the final factor is a magnetic correction, $B_\mu \equiv \sqrt{4\pi\mu} \approx 4 \times 10^{15} \rho_{14}^{0.4}$ Gauss and $\rho_{14} \approx 1$ is the density in the deep crust in units of $10^{14}$ g cm$^{-3}$.

The 92.5 Hz QPO and the weaker feature at 30.4 Hz discovered in SGR 1806-20 are suggested to be due to the $\nu(_7t_0)$ and $\nu(_2t_0)$ modes, respectively and the detection of the 626.5 Hz QPO was modeled as the $n=1$ toroidal mode where the energy required to excite $n>0$ modes is orders or magnitude larger than the energy required to excite an $n=0$ mode which reflects the intensity of the flare. Excitement of higher order $l$ modes effect was seen in the spectrum of Earth modes excited by the 2004 Sumatra-Andaman earthquake [7]. In SGR 1900+14 a sequence of modes of different $l$ is plausible in which the frequencies are due to the $l=4$ (53 Hz), $l=7$ (84 Hz), and $l=13$ (155 Hz) modes with fundamental toroidal mode frequency $\nu(_2t_0) \approx 28$ Hz, lower than that for 1806 yet consistent with the 28 Hz QPO detected in 1900.

We discuss our findings in context of the toroidal mode frequencies where the QPO candidates in 1806 at 84 and 104 correspond to $\nu(_6t_0)$ and $\nu(_8t_0)$ respectively. While the high frequency feature we report in SGR 1806-20 at 648 Hz may be interpreted in terms of $n=1$ toroidal crust mode. Another recent interpretation of the QPOs in the frequency range from 18 to 1840 Hz could be modeled and well accounted for using the standing sausage mode oscillations of flux tubes in the magnetar corona [22]. It is assumed that part of the plasma ejected during the giant flares is trapped by the magnetic fields and then forms magnetic flux tube structures similar to what is seen in the solar corona. Unfortunately the high frequency candidates we report at 1096, 1230, 2785 and 3690 Hz in SGR 1806-20 are difficult to explain in the tube oscillation model. For these extremely high Eigen frequencies, the $n > 1$ and torsional mode shear vibrations in the neutron star crust might be a more plausible interpretation [13].

The short rise time of the flares analyzed in this work may introduce spurious power peaks even if there are no actual periodic signals present which may raise some skepticism about the source of these oscillations in the magnetar bursts. We employed Monte Carlo simulations technique on the features reported in this paper and our results are forthcoming.

## ACKNOWLEDGMENTS

The authors would like to thank Anna Watts for her insightful comments and suggestions and Craig Markwardt for the useful discussions. One of us (A.M) is grateful for the valuable discussions with Kent Wood and Christopher Thompson during the Neutron Stars & Gamma Ray Bursts Conference 2009 held in Cairo and Alexandria.